%% file: main.tex
\documentclass[10pt,twocolumn,letterpaper]{article}

\usepackage[pagenumbers]{cvpr} 

\input{preamble}

\definecolor{cvprblue}{rgb}{0.21,0.49,0.74}
\usepackage[pagebackref,breaklinks,colorlinks,citecolor=cvprblue]{hyperref}

\usepackage{booktabs} 
\usepackage{amsmath}
\usepackage{colortbl}
\usepackage{amsthm}
\usepackage{gensymb}
\usepackage{multirow}
\usepackage{enumitem} 
\usepackage{mathtools}

\usepackage{pifont}
\newcommand{\cmark}{\textcolor{green}{\ding{51}}}%
\newcommand{\xmark}{\textcolor{red}{\ding{55}}}%

\newcommand{\norm}[1]{\left\lVert#1\right\rVert}
\def\MethodName{VNCA}

\title{Volumetric Temporal Texture Synthesis for Smoke Stylization \\ using Neural Cellular Automata}

\author{Dongqing Wang \quad Ehsan Pajouheshgar \quad Yitao Xu \quad Tong Zhang \quad Sabine Süsstrunk\\
School of Computer and Communication Sciences, EPFL \\
{\tt\small \{dongqing.wang, tong.zhang, alaa.abboud, sabine.susstrunk\}@epfl.ch}
} 

\begin{document}

\input{teaser}
\twocolumn[{%
\maketitle%
\figfirstpagefigure%
}]

\input{sec/0-abstract}
\input{sec/1-intro}
\input{sec/2-related}

\input{sec/3-method}

\input{sec/4-exp}

\input{sec/5-mesh}

\input{sec/6-concl}

{
    \small
    \bibliographystyle{ieeenat_fullname}
    \bibliography{main}
}

\begin{figure*}
  \centering
  \includegraphics[width=0.8\textwidth]{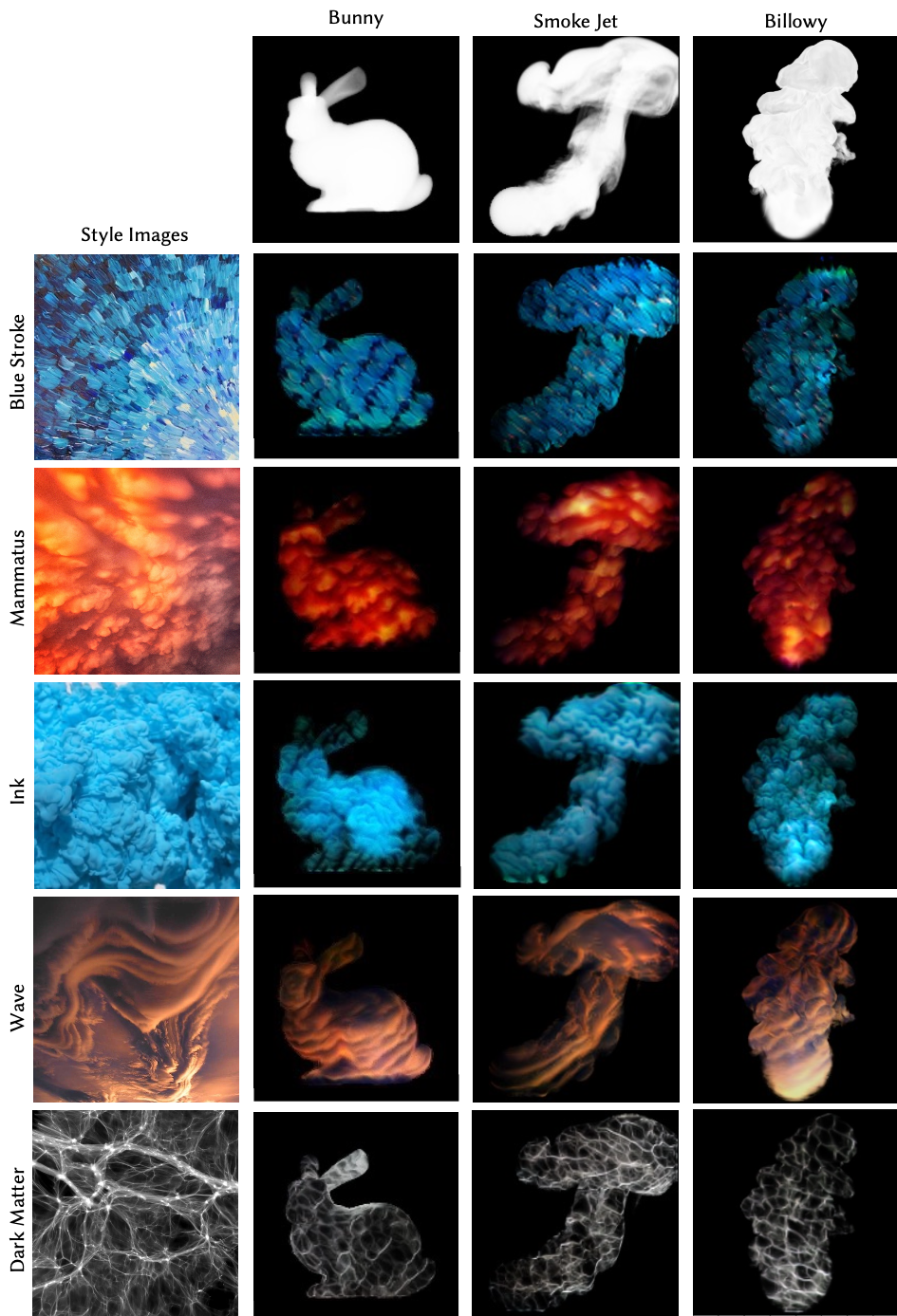}
  \caption{\textbf{Stylization on Smoke Simulations. } VNCA can both synthesize RGB color and modify density distribution for a stylization consistent with the input style image.  } 
  \label{fig:res}
\end{figure*}

\begin{figure*}
  \centering
  \includegraphics[width=0.85\textwidth]{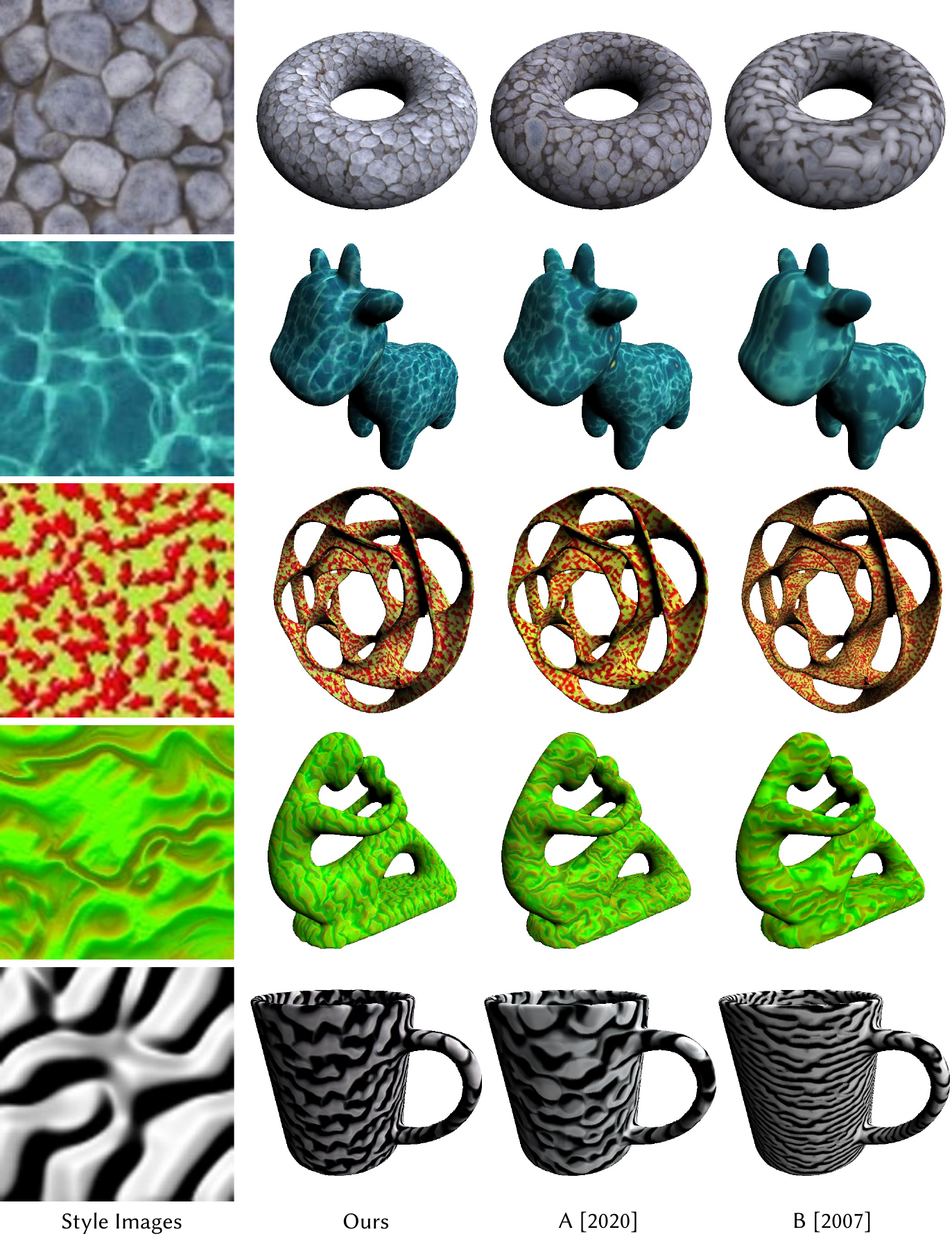}
  \caption{\textbf{Stylization on Meshes. } VNCA can be generalized and applied to mesh texturing. A. \citet{on-demand-solid-texture}, B. \citet{kopf2007solid}. Our method synthesizes mesh stylization with better consistency with the style image and fine-level details. }
  \label{fig:mesh}
\end{figure*}

\end{document}

%% file: preamble.tex
\usepackage[dvipsnames]{xcolor}


%% file: teaser.tex
\newcommand{\figfirstpagefigure}{
  \begin{center}
  \includegraphics[width=\textwidth]{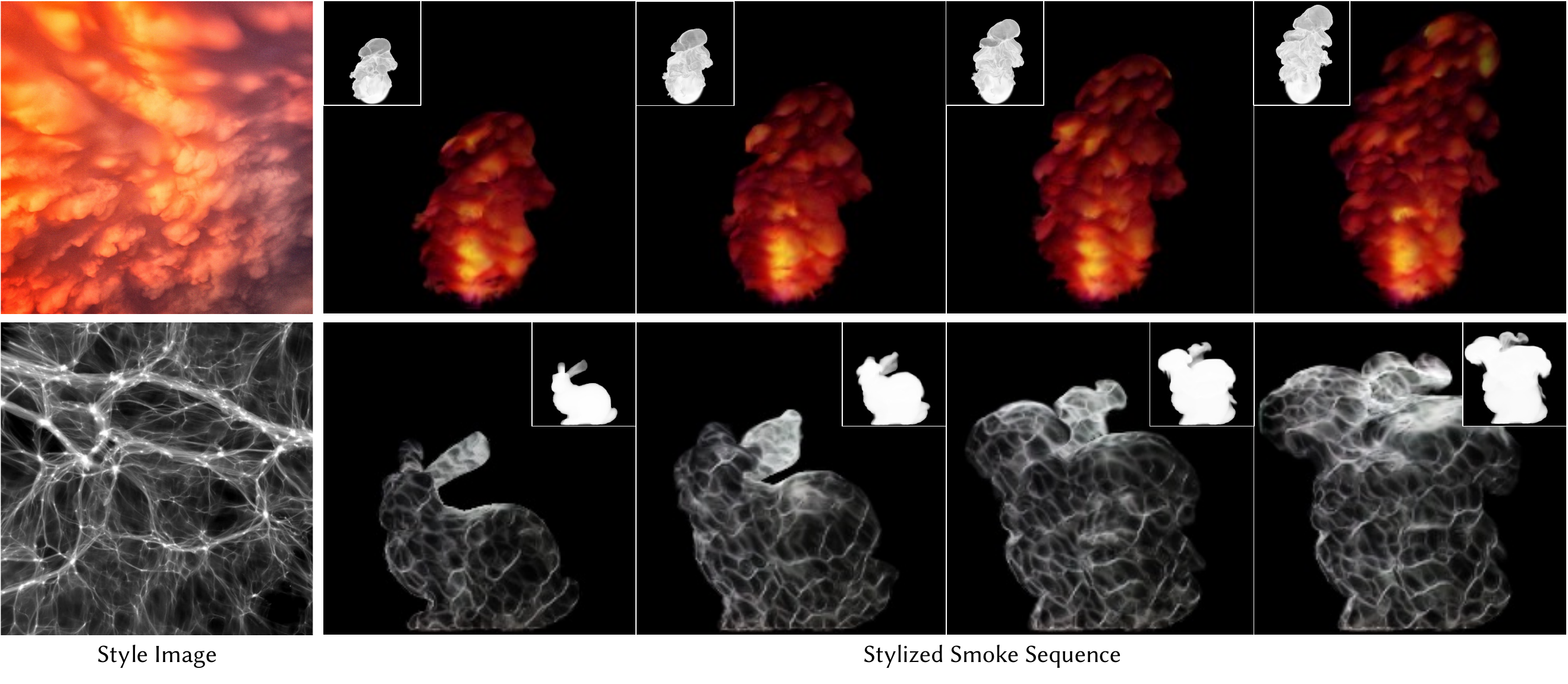}
  \captionof{figure}{Given a simulated 3D smoke sequence and a reference style image, we train our Volumetric Neural Cellular Automata (VNCA) model on a single density frame of the simulation to synthesize a dynamic texture volume. For inference, our trained model can stylize the whole smoke sequence in real-time. We recommend the readers check our images on screen for optimized visual quality.}
  \label{fig:teaser}
    \end{center}
}

%% file: sec/0-abstract.tex
\begin{abstract}
Artistic stylization of 3D volumetric smoke data is still a challenge in computer graphics due to the difficulty of ensuring spatiotemporal consistency given a reference style image, and that within reasonable time and computational resources.
In this work, we introduce \textbf{V}olumetric \textbf{N}eural \textbf{C}ellular \textbf{A}utomata (\MethodName{}), a novel model for efficient volumetric style transfer that synthesizes, in real-time, multi-view consistent stylizing features on the target smoke with temporally coherent transitions between stylized simulation frames.
 \MethodName{} synthesizes a 3D texture volume with color and density stylization and dynamically aligns this volume with the intricate motion patterns of the smoke simulation under the Eulerian framework. 
Our approach replaces the explicit fluid advection modeling and the inter-frame smoothing terms with the self-emerging motion of the underlying cellular automaton, thus reducing the training time by over an order of magnitude. 
 Beyond smoke simulations, we demonstrate the versatility of our approach by showcasing its applicability to mesh stylization.

\end{abstract}

%% file: sec/1-intro.tex
\section{Introduction}
Creating realistic and artistically expressive 3D volumetric smoke is a complex task in computer graphics, with applications in film productions. 
The nonlinearity of numerous physical factors within the fluid motion, described by the Navier–Stokes equations~\cite{bridson2015fluid}, lead to the inherent turbulence and ever-changing patterns in smoke simulations. 
To assist artists in achieving creative smoke stylization, there is a need for an intuitive way to input style references and get stylization results quickly. 
It is challenging to preserve the \textbf{spatiotemporal} consistency of the stylized volumetric smoke: maintaining spatial and temporal consistency across multiview and consecutive frames, together with having a stylization that matches the style image and preserving an overall plausible fluid motion. 
Additionally, high-resolution smoke simulations require a lightweight rendering algorithm to manage memory efficiently during training.

\input{sec/table-related}

Existing work for volumetric smoke stylization with a given style image can be classified into two types, see \Cref{tab:baseline}. 
\textit{Optimization-based} volumetric Neural Style Transfer (NST) techniques model 3D fluid through either Eulerian~\cite{kim19c} or Lagrangian~\cite{kimlnst} framework, iteratively stylizing the smoke simulation with loss functions similar to image style transfer~\cite{gatys2016image}. 
While these methods achieve temporal coherence by introducing an inter-frame smoothing loss, their computation costs scale up excessively and become unpractical for multi-view optimizations~\cite{kanyuk2023singed}.  
Other approaches~\cite{guo2021volumetric, aurand2022efficient} train \textit{feed-forward} 3D convolutional neural networks (CNN), using the original volume and style image as input to directly output the stylized volume.
These methods can produce multi-view consistent stylization. 
However, as a consequence of using large neural networks with millions of parameters, the training time for density-only stylization may take up to 20 hours, and 70 hours for colored albedo volumes.
\Cref{tab:baseline} summarizes the characteristics of the two categories of methods. 

In this work, we provide a new perspective on volumetric stylization by synthesizing a 3D texture volume that is inherently dynamic and continuously aligns with input smoke motion.
We propose Volumetric Neural Cellular Automata (\MethodName{}), a novel method that efficiently achieves high spatiotemporal consistency for stylizing smoke simulation given a reference style image. 
Our approach is reminiscent of \textit{Solid Texture} methods for mesh texturing \cite{on-demand-solid-texture, kopf2007solid, solid_textures}, with the key distinction that our synthesized texture volume is dynamic and iteratively updated, matching the motion of the smoke simulation.

\MethodName{} is parameterized by a small \textit{recurrent} neural network\footnote{VNCA update rule is applied over different timesteps to update cell states. The cell states in VNCA resemble the hidden state in RNNs and the VNCA update rule resembles the time-invariant neural network of RNNs. } and is fast to train.  
It is trained on a single frame of the density sequence, and yet it can generalize to unseen density frames during inference for stylization in \textit{real-time}. 
We utilize the inherent spontaneous moving patterns of NCA-generated textures and match the motion of the smoke simulation to create a visually plausible fluid stylization. 
Since our synthesized 3D texture volume is naturally dynamic, we do not need to explicitly model fluid advection, nor to apply interframe smoothing terms to enforce temporal coherence. 
These advantages, coupled with the strong generalization ability of our NCA model, allow us to achieve a speedup in training of over an order of magnitude while creating high-quality smoke stylization.
Moreover, we show that our method can be adapted to mesh stylization by utilizing a similar training strategy as Solid Texturing methods, achieving results on par with existing methods.

\noindent To summarize, our contributions are as follows:
\begin{itemize}[leftmargin=*]
    \item We propose \MethodName{}, a novel method to stylize 3D smoke simulations in real-time. VNCA generalizes to the whole simulation sequence while trained on a single frame. 
    \item Our stylization for the 3D smoke simulation is spatiotemporally consistent w.r.t the input reference image. 
    \item \MethodName{} speeds up training by over an order of magnitude. 
    \item \MethodName{} synthesizes texture volumes for mesh stylization. 
\end{itemize}

%% file: sec/table-related.tex
\newcommand\RotText[1]{\rotatebox{90}{\parbox{2cm}{\centering#1}}}

\begin{table}[t]
\scalebox{0.9}{
\begin{tabular}{c|cccccccc}
\begin{tabular}[c]{@{}c@{}}\textbf{Methods}\end{tabular} & 
\begin{tabular}[c]{@{}c@{}}{\textbf{A}}\end{tabular} &
\begin{tabular}[c]{@{}c@{}}{\textbf{B}}\end{tabular} &
\begin{tabular}[c]{@{}c@{}}{\textbf{C}}\end{tabular} & 
\begin{tabular}[c]{@{}c@{}}{\textbf{D}}\end{tabular} & 
\begin{tabular}[c]{@{}c@{}}{\textbf{E}}\end{tabular} &
\begin{tabular}[c]{@{}c@{}}{\textbf{F}}\end{tabular} &
\begin{tabular}[c]{@{}c@{}}{\textbf{G}}\end{tabular} &
\textbf{Type}  \\
\toprule
\citet{kim19c}           & \xmark  & \cmark  & \xmark  & \xmark   & \xmark & \cmark & \xmark &
\multirow{4}{*}{\rotatebox{90}{\parbox{1.3cm}{\centering \textbf{\hspace{3pt} \small{Optim.} }}}} \\

\citet{kimlnst}           & \xmark  & \cmark  & \xmark  & \xmark   & \xmark & \cmark & \xmark &  \\

\citet{aurand2022efficient} & \xmark  & \cmark  & \xmark  & \xmark   & \cmark & \cmark & \xmark &  \\

\midrule
\citet{guo2021volumetric}           & \cmark  & \cmark &  \cmark  & \xmark   & \cmark & \xmark & \xmark &  
\multirow{3}{*}{\rotatebox{90}{\parbox{1.1cm}{\centering \textbf{\hspace{2pt} \small{Feed \\ Forward}}}}} \\

\citet{aurand2022efficient} & \cmark  & \xmark &  \xmark  & \xmark   & \cmark & \xmark & \xmark &\\
\textbf{VNCA}  & \cmark  & \cmark &  \cmark  & \cmark   & \cmark & \xmark & \cmark &  \\
\bottomrule
\end{tabular}
}
\caption{\textbf{Relevant methods. } \textbf{A.} Models multiview stylization; \textbf{B.} Models temporal consistency between frames; \textbf{C.} Trained model generalizes to other smoke data; \textbf{D.} Stylizes both color and density in 3D; \textbf{E.} Inference time $\leq$ 1m/f; \textbf{F.} Training GPU Memory $\leq$ 12GB; \textbf{G.} Generalizes to mesh texturing. } 
\label{tab:baseline}
\end{table}

%% file: sec/2-related.tex
\section{Related Works}

\subsection{Texture Synthesis with Neural Cellular Automata. }
Cellular Automata (CA) consists of a regular grid of cells that are updated based on their neighboring cell states. 
One famous example is \textit{Game of Life}~\cite{gardner1970fantastic}. 
All cells follow the same update rule asynchronously. 
\citet{gilpin2019cellular} show that the update rule of CA can be parameterized with neural nets, resulting in Neural Cellular Automata (NCA). 
\citet{niklasson2021selforganising} first use NCA to synthesize 2D textures with self-emerging, temporally varying motion with a static image exemplar as input. 
\citet{diff-program-rdsystem} train a 2D reaction-diffusion system that can be extended to 3D at inference time. 
However, it cannot control the synthesized motion due to Laplacian's isotropic nature. 
\citet{pajouheshgar2023dynca} modify the NCA model for controllable 2D dynamic texture through a pre-trained optical-flow network~\cite{tesfaldet2018two} and vector field supervision.
Despite the improvements to NCA's performance and generative abilities for texture synthesis, none use the self-emerging motion to its full advantage.

Here, we introduce a variation of NCA on 3D volume grids.
Our \MethodName{} exploits the movement that naturally emerges from NCAs post-training. These motion patterns are suitable replacements for the costly advection in fluid simulation and stylization. 
Additionally, VNCA conditioned on a specific density field during training can stylize unseen density fields at inference time.

\subsection{Solid Texture Synthesis.}
Solid texturing, introduced by \citet{solid_textures} and \citet{perlin_noise}, enables texturing meshes without UV maps. 
While the early solid texturing methods \cite{solid_textures, perlin_noise} require hand-designed functions for each texture, \citet{kopf2007solid} propose a non-parametric, optimization-based algorithm for an explicit solid texture volume from a 2D exemplar texture. 
Several works leverage neural nets to create a solid texture block and optimize its parameters based on the target exemplar texture \cite{gramgan, on-demand-solid-texture, neural-texture-space}. 
These methods evaluate the 2D slices of the synthesized solid texture using a texture loss \cite{gatys2015texture} for optimization. 

Our approach to the stylization of smoke simulation sequences can be seen as spatiotemporal modeling of volumetric texture. 
It assigns texture to a bounded region around the smoke sequence, similar to the process of solid texture synthesis.
The synthesized texture volume is naturally dynamic and is trained to align with the input density sequence's motion from fluid advection governed by the Navier-Stokes Equations~\cite{bridson2015fluid}.

\subsection{Volumetric Neural Style Transfer.}
Prior works on stylizing volumetric smoke data guided by exemplar images can be categorized into optimization-based methods and feed-forward methods. 
\textbf{Optimization based} algorithms \cite{kim19c, kimlnst, aurand2022efficient} extract style features using the pre-trained VGG \cite{vgg16} from both the exemplar image and the rendered smoke appearance, and iteratively optimize the density using style-based loss functions \cite{gatys2016image}. 
Transport-based style transfer (TNST)~\cite{kim19c} synthesizes complex stylization on Eulerian-based smoke, but suffers from lengthy optimization and temporal inconsistency due to its costly temporal smoothing scheme. 
The Eulerian framework observes fluid flow at fixed points in space over time.
To resolve these two drawbacks, \citet{kimlnst} reformulate TNST in a Lagrangian framework at the cost of protracted conversion time from smoke density to a particle-based representation.
Meanwhile, ~\citet{aurand2022efficient} enhance the performance by using lighter advection and temporal smoothing algorithms. 
However, These methods still suffer from the slow explicit modeling of advection, and the optimization time scales up with the number of input frames and the number of views used for training. 
Finally, these methods do not support color style transfer on 3D volume data. 
\textbf{Feed-forward} methods~\cite{guo2021volumetric, aurand2022efficient} train 3D CNNs to directly output stylized smoke density, bypassing iterative optimizations. 
These methods ensure multi-view style consistency by directly processing 3D data.
\citet{guo2021volumetric} use a 3D volume autoencoder and trains a model for arbitrary style images while incorporating a regularization for temporal consistency. 
However, it requires 70+ hours of training, and in practice, not all images can produce multiview consistent stylization for volume data.
\citet{aurand2022efficient} train separate networks for each style and smoke pair to improve the multi-view consistency, taking 18 hours to train one network. 
However, it has no explicit control on temporal coherence except implicit translational equivariance. The use of tiling due to the memory requirement to output the full stylized smoke at inference further harms temporal consistency.  

Our volume stylization technique presents a novel interpretation by framing volumetric stylization as a texturing problem, synthesizing 3D textures that are naturally dynamic. 
VNCA significantly speeds up training compared to the prior works. At the same time, our method maintains multiview stylization coherence and temporal smoothness across each stylized frame of 3D smoke data. 
Inspired by advances in 2D dynamic texture~\cite{pajouheshgar2023dynca}, we imitate fluid advection by controlling the self-emerging motion pattern of NCA to preserve a natural temporal consistency.

\begin{figure}
  \centering
  \includegraphics[width=\linewidth]{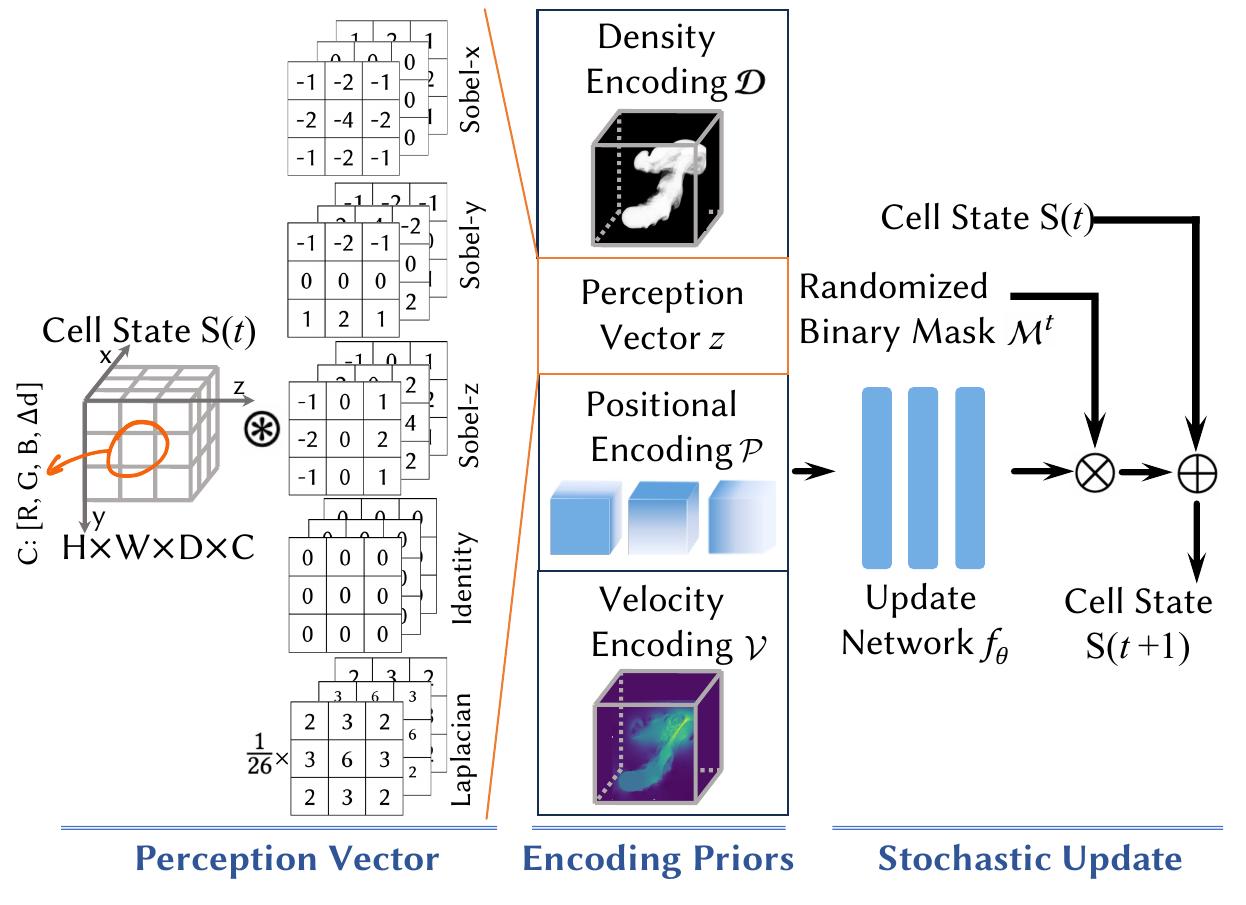}
  \caption{\textbf{Update Rule.} We use the update rule to determine the cell state at the next time step. We obtain the perception vector with neighboring cell state and concatenate density and positional encoding as priors. The stochastic update processes this vector for an update to each cell.}
  \label{fig:update}
\end{figure}

%% file: sec/3-method.tex
\section{Method}
\subsection{Background: Neural Cellular Automata}
NCA operates on a 2D grid of cells with size $H\times W$, and each cell stores its state information in a vector with $C$ channels. 
The first 3 channels are interpreted as R, G, and B values, and the cell states are initialized with zero values. 
At each time step, the cell state updates once. 
$\mathbf{S}^t \in \mathbb{R}^{H\times W\times C}$ denotes the \textit{cell states} at time $t$ and $s^t_{ij} \in \mathbb{R}^C$ represents the state of the cell at position ($i, j$).  
The \textit{update rule} is a trainable PDE that determines how $s^t_{ij}$ changes over time depending on the cell state and its spatial gradients and Laplacian. 
All cells share the same local update rule which comprises \textit{two steps}. 
First, each cell state convolves a set of filters over its surrounding cells to form a \textit{perception vector} containing neighborhood information. 
Then, the perception vector is passed through two Fully Connected layers and a random binary mask to determine the new cell state.

NCA is notable for its low number of trainable parameters, self-emerged motion over synthesized texture, and strong generalizability.  
A trained NCA exhibits temporal coherence and provides a manifold of solutions aligned with the style image, which can be perceived as motion. 
Its generalizability to unseen situations is given by the shared update rule. 
We exploit the emerging motion to simulate visually plausible stylized fluid advection, and use its generalizability to stylize unseen density fields at inference.

\subsection{Volumetric NCA Representation} 
\label{sec:vnca}
We define the Volumetric NCA grids on a 3D voxel-based representation. 
To avoid confusion, we refer to the voxels in VNCA as ``cell'', and the voxels in the input density field as ``voxels''. 
Note that the time step $t$ in NCA updates is not one-to-one mapped to the frames of the input smoke simulation. 
As detailed in \Cref{sec:motion}, we map $N$ steps of cell state update to two adjacent density frames for an approximated fluid motion at inference time.  

Our VNCA model is applied to each density frame $d: \mathbb{R}^3 \rightarrow \mathbb{R}$ for color and density stylization, instead of predicting the next density frame. 
We construct our VNCA through a voxelized cube of resolution $H\times W\times D$. 
The cell state at time $t$ is thus denoted as $\mathbf{S}^t \in \mathbb{R}^{H\times W\times D \times C}$, with $s^t_{ijk} \in \mathbb{R}^C$ as a vector denoting the $C$-channel cell state at cell ($i$, $j$, $k$).
The first 3 channels represent R, G, and B values for the corresponding voxel in the input density field, respectively, and the 4th channel represents $\Delta d$ as the amount of density to be applied to the corresponding voxel for stylization, detailed in \Cref{sec:dvr}.
All channels are initialized with 0s.

\subsection{Update Rule}
\label{sec:update}
VNCA's update rule defines a partial differential equation (PDE) over the cell states:
\begin{equation}
    \mathbf{S}^{t+\Delta t} = \mathbf{S}^t + \frac{\partial \mathbf{S}^t}{\partial t}\Delta t,
\end{equation}
We solve this PDE by discretizing the 3D space using a voxelized grid and applying the update rule with fixed-sized time steps. \Cref{fig:update} illustrates the update rule along with the VNCA cell state.

\medskip
\noindent{\textbf{Perception vector.}}
To perceive information from all neighborhood cells in the context of a 3D voxel, we apply 3D convolutions with a set of filters to the cell states to obtain the discretized gradient and Laplacian at the cell's location and form its perception vector. 
The filters are the Sobel operators and the 27-point variant of the discrete Laplace operator according to ~\citet{o2006family} as filters. 
Their weights are frozen during training. 
The perception vector $z_{ijk} \in \mathbb{R}^{5C}$ for $s_{ijk}$ is given by the concatenation of the cell state, the discretized gradient and laplacian: 
\begin{equation}
    z_{ijk} = s_{ijk} \;\Vert \; (\partial_\text{x}\mathbf{S}|_{ijk}) \; \Vert \;(\partial_\text{y}\mathbf{S}|_{ijk})\; \Vert \;(\partial_\text{z}\mathbf{S}|_{ijk})\;\Vert \;(\triangledown^2\mathbf{S}|_{ijk}). 
\end{equation}

\medskip
\noindent{\textbf{Encoding Priors.}}
The perception vector is designed to update the cell states towards an aligned style and visually plausible motion. 
Therefore, the perception vector should be aware of essential information about the stylized objective.
We thus encode the 1. spatial location of each cell, 2. the distribution of the input smoke density field, and 3. the velocity provided by the input smoke data at the current location and time step as three priors to the perception vector, so cell updates take these priors into account. 

Specifically, we apply a 3-dimensional \textbf{positional encoding}~\cite{pajouheshgar2023dynca}, a \textbf{density encoding}, and a \textbf{velocity encoding} as shown in \Cref{fig:update}. 
Our positional encoding tensor has three channels and is defined as: 
 $\mathcal{P}_{ijk} = \left[\frac{2i + 1}{H} - 1, \frac{2j + 1}{W} - 1, \frac{2k + 1}{D} -1 \right]. $
Using positional encoding improves the stability and consistency of the synthesized motion.  
For density encoding, we append the 1-channel input density field $\mathcal{D}$ of the current smoke frame in training to the end of the perception vector to allow VNCA to be aware of the structure of input density.
Density encoding encourages style features that closely match the input reference, as opposed to a "perforated" appearance. 
The velocity encoding $\mathcal{V}$ is meant to interpolate the update for the desired advected density field. 
It provides spatial inductive bias to help VNCA better match the advection of the input sequence, and produce visually plausible motion. We provide ablation studies in \Cref{sec:abl}

\medskip
\noindent{\textbf{Stochastic Update.}}
We pass the encoded perception vector through a two-layer Multi-Layer Perceptron (MLP) $f_\theta$ with a ReLU activation to determine the updated state of each cell. Similar to the original NCA model, each cell update follows a geometric process~\cite{geometric-process} independently from the other cells. 
This asynchronous cell update is implemented by applying a randomized binary mask $\mathcal{M}^t$ to MLP output:
\begin{equation}
    \frac{\partial s_{ijk}}{\partial t} = \mathcal{M}^t(f_\theta(z_{ijk} \; \Vert \; \mathcal{P}_{ijk}\; \Vert\; \mathcal{D}_{ijk})). 
\end{equation}
Each voxel in the random binary mask $\mathcal{M}$ is a Bernoulli random variable with probability $p=0.5$.
This asynchronous cell update scheme improves the robustness and stability of VNCA and allows stylizing longer frame sequences.

\subsection{Training}
\label{sec:opt}
\begin{figure}
  \centering
  \includegraphics[width=1.03\linewidth]{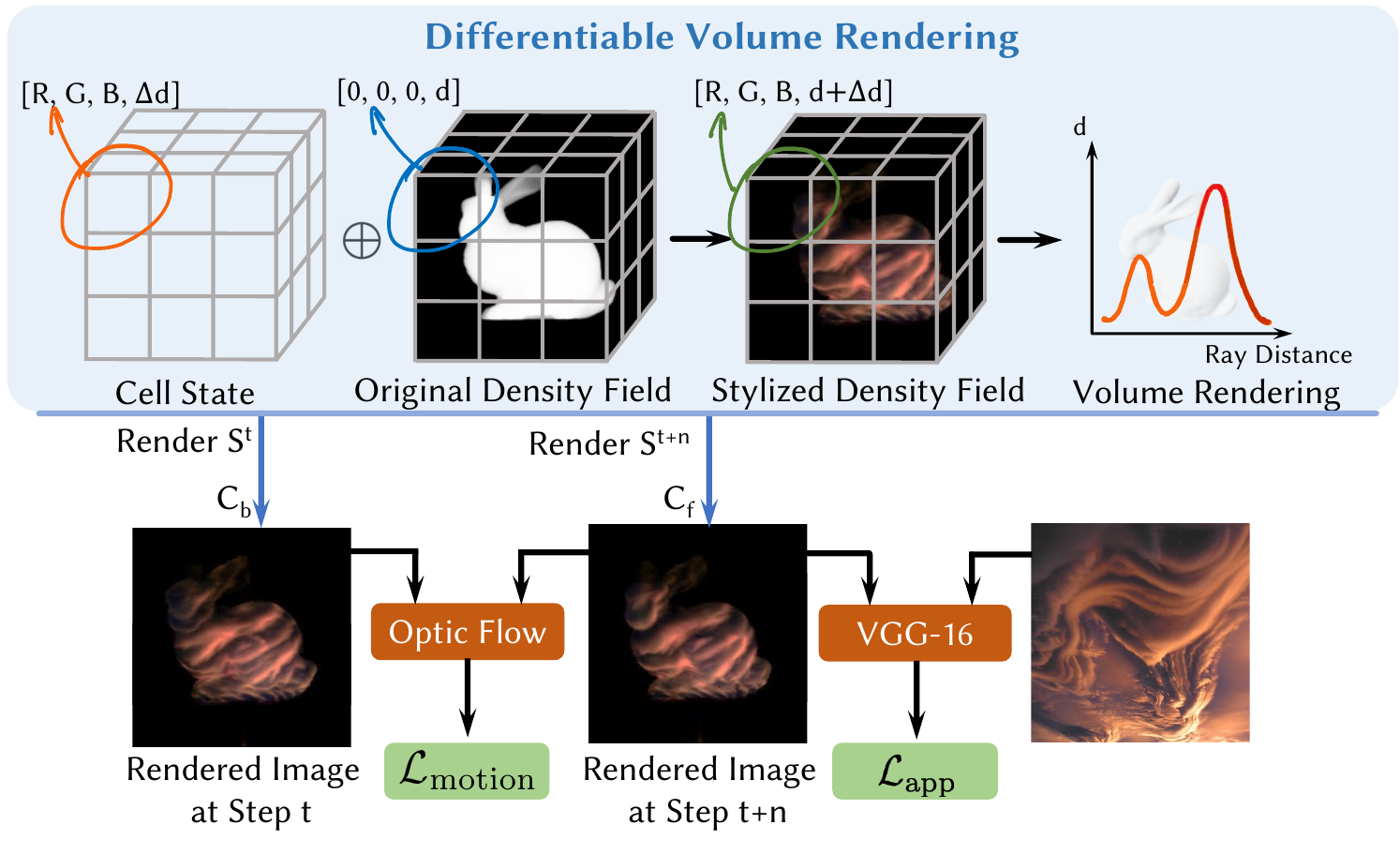}
  \caption{\textbf{Training.} At training, we use a single input frame for the \textit{Density Encoding} shown in Fig. 2. At each epoch, we apply the VNCA update rule for $n$ steps and record the cell state before and after the update. We then render the smoke stylized with cell state, using Eq.7, before and after the VNCA updates to get $P_b$ and $P_f$. We use $P_f$ to match the style image and extract motion between $P_b$ and $P_f$ to align with the fluid motion. }
  \label{fig:optim}
\end{figure}

We represent the volumetric temporal texture through a trained VNCA. 
This allows for real-time stylization that aligns with both the reference image and the motion of the input smoke sequence. 
\Cref{fig:optim} shows the training pipeline.

\medskip
\noindent{\textbf{Differentiable Volume Rendering.}} 
\label{sec:dvr}
We represent the smoke data under the Eulerian framework. 
To make training more efficient, we implement a lightweight differentiable volume rendering algorithm that measures the radiance for each ray $\mathbf{r}$ transmitted through an inhomogeneous participating medium~\cite{fong2017production}. 
The original method calculates transmittance $\tau$ and a grayscale pixel intensity $I_{ij}$ corresponding to occupancy by accumulating the densities along the ray through equidistant sampling:
\begin{equation}
    \tau(k, \mathbf{r}) = e^{-\gamma\sum_k^D d(\mathbf{r}, k)}, \quad I_{ij} = \sum_{k=0}^D d(i,j,k)\tau(k, \mathbf{r}_{ij}),
\end{equation}
where k is the index of the voxel into the density dimension; $\mathbf{r}_{ij}$ is a ray vector traced from pixel ($i$, $j$) in the image space to the normal direction of our orthographic camera; d($\cdot$) is the density value in the volume grid; $\gamma$ is the transmittance absorption constant, which describes the rate at which the volume absorbs light. 

Our implementation is based on this algorithm and incorporates colorization into the volume with a NeRF-styled volume renderer from ~\citet{mildenhall2020nerf}. 
We assume that voxel ($i$, $j$, $k$) not only contains a density value $d(i,j,k)$ but also emits radiance $c(i,j,k)$ whose RGB value is assigned by the first three channels of the VNCA cell states. 
The fourth channel of the VNCA cell states $\Delta d(i,j,k)\in [-0.5, 0.5] $ stores the residual update on the input density. 
The RGB color of the pixel $P_{ij}$ is therefore defined as:
\begin{equation}
    \tau(k, \mathbf{r}) = \exp{-\gamma\sum_k^D \max(d(\mathbf{r}, k)( 1 + \Delta d(\mathbf{r}, k)), 0)},
\end{equation}
\begin{equation}
\label{eq:render}
P_{ij} = \sum^D_{k=0} c(k)\max(d(\mathbf{r}_{ij},k)(1 + \Delta d(\mathbf{r}_{ij},k)), 0)\tau(k, \mathbf{r}_{ij}). 
\end{equation}
In practice, we choose $C = 12$ channels for the state of each VNCA cell. The remaining 8 channels are hidden channels that do not participate in the loss computation. 
We empirically observe that including hidden channels in the cell state stabilizes the training and increases stylization quality.

\medskip
\noindent{\textbf{Appearance Supervision.}} 
After obtaining the rendered image of the stylized smoke $P_f$, we use a pre-trained VGG16 network~\cite{vgg16} to extract the style information using deep feature representations~\cite{gatys2016image}. 
The local cell communication in the update rule of VNCA is shared by all cells and the rendered 2D image involves contribution from all cells. 
Given this inductive bias, using only a single rendered view at each epoch is sufficient for training VNCA to a converged solution that matches the reference image from all views,  contributing to our faster training compared to the baseline methods. 
Note that we change the camera viewpoint across training epochs. 

Our appearance loss is based on the style loss proposed by \citet{kolkin2019style} and consists of style-matching and moment-matching terms. 
We extract two sets of VGG16 feature maps $F^r$ and $F^\psi$ of size $C'\times H' \times W'$ from the rendered image and the target style image, respectively. $C'$ and $H' \times W'$ represent the number of channels and the spatial dimensions of the VGG16 feature maps respectively.
These feature maps are then flattened along the spatial dimensions to obtain the feature sets $A^r$ and $A^\psi$.
We measure the style distance $D_c$ between features through the minimum cost of transporting from one set to another using cosine distance:
\begin{equation}
d_{c}(x, y) = 1 - \frac{x \cdot y}{\norm{x}_2 \norm{y}_2},
\end{equation}
\begin{equation}
D_c(A^r, A^\psi) = \frac{1}{H' \times W'}\sum_i\min_j d_{c}(A^r_i, A^\psi_j).
\end{equation}
The subscript is the feature vector index in the feature set.

To encourage density modifications in the stylized output, we render the stylized smoke in \textit{grayscale}. 
We then extract deep feature maps $F_{\textup{gray}}^r$ and $F_{\textup{gray}}^\psi$ and feature sets $A_{\textup{gray}}^r$ and $A_{\textup{gray}}^\psi$ from the rendered image and the grayscale stylized image as described above, and similarly define a distance for density pattern matching.  

Our final style-matching term $\mathcal{L}_\text{style}$ combines the style distance between the grayscale and colored images:
\begin{equation}
    \begin{aligned}
        \overleftrightarrow{D_c} (X, Y) = \max(D_c(X, Y), D_c(Y, X)) \\
        \mathcal{L}_\text{style} = \overleftrightarrow{D_c}(A^r, A^\psi)+ \overleftrightarrow{D_c}(A_{\textup{gray}}^r, {A}_{\textup{gray}}^\psi).
    \end{aligned}
\end{equation}

We additionally include a moment-matching term $\mathcal{L}_\text{moment}$ to align the magnitudes between colored features $A^r$ and $A^\psi$, and between grayscale features $A_{\textup{gray}}^r$ and $A_{\textup{gray}}^\psi$:
\begin{equation}
    \begin{aligned}
        D_k(X, Y) = \frac{1}{C'} \norm{\mu_{X} - \mu_{Y}}_1 +  \frac{1}{C'^2} \norm{\Sigma_{X} - \Sigma_{Y}}_1 \\
        \mathcal{L}_\text{moment} = D_k(A^r, A^\psi) + D_k({A}_{\textup{gray}}^r, {A}_{\textup{gray}}^\psi). 
    \end{aligned}
\end{equation}
where $\mu_{X}$ and $\Sigma_{X}$ denote the mean and covariance of the feature set $X$. 
Our overall appearance loss is the sum of the style-matching and moment-matching terms:
\begin{equation}
     \mathcal{L}_\text{app} = \mathcal{L}_\text{style} + \mathcal{L}_\text{moment}.
\end{equation}
In our experiments, we use deep features at different scales and sum our appearance loss over 5 layers of VGG16.

\begin{figure}
  \centering
  \includegraphics[width=0.9\linewidth]{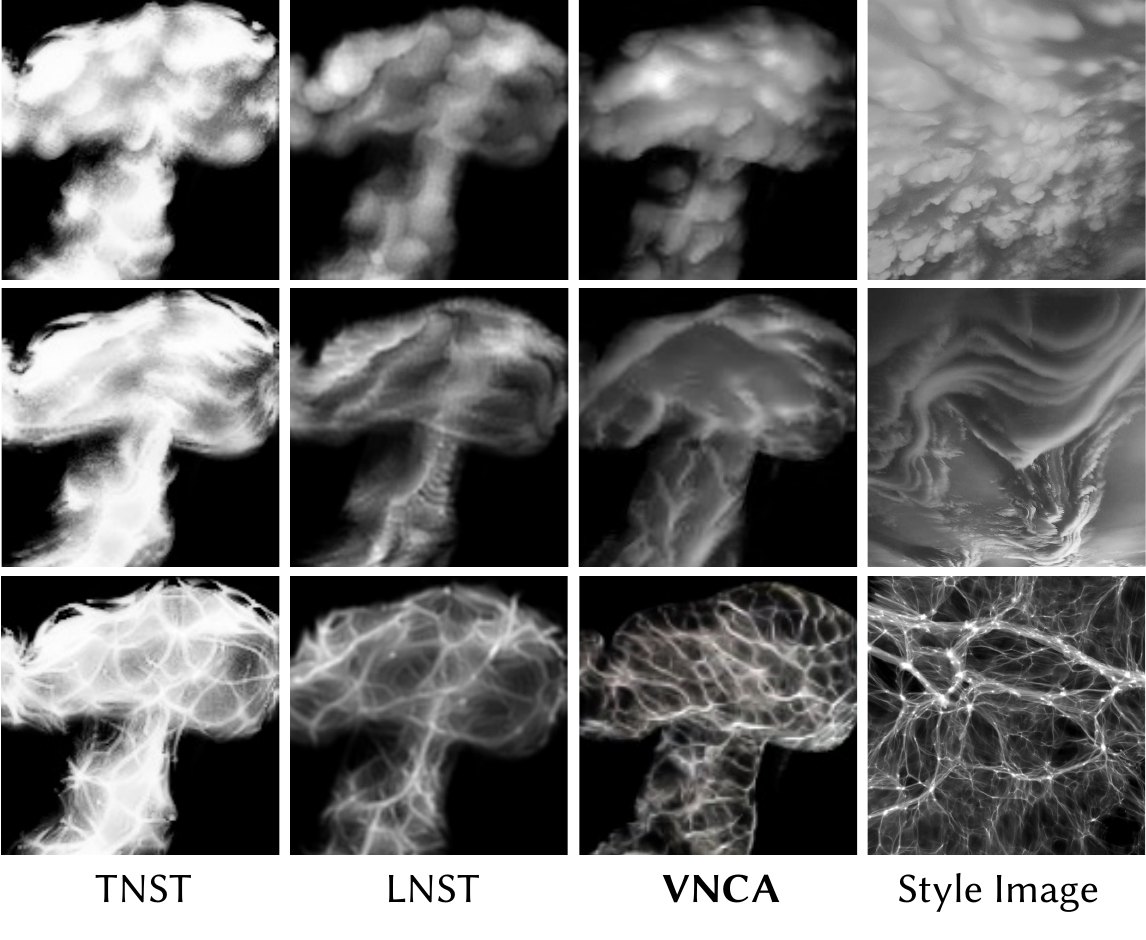}
  \caption{\textbf{Qualitative Stylization Comparison with TNST and LNST.} VNCA synthesizes stylization that closely matches the reference image, with superior quality than TNST and LNST indicated by our user study.}
  \label{fig:comp}
\end{figure}

\medskip
\noindent{\textbf{Flow-guided Motion Supervision. }}
\label{sec:motion}
Instead of explicitly simulating transport, we supervise the emerged motion of the synthesized temporal texture with the input velocity field to align with the smoke movement.
Since a pre-trained model for estimating 3D motion vectors does not exist, it is costly and unfeasible to directly compare the 3D smoke motion and our volumetric temporal texture. 

We project the velocity field of input smoke data onto a 2D camera plane with camera pose $\Phi$ and denote the projection as $\mathcal{J}^\psi \in \mathbb{R}^{H\times W\times 2}$ to supervise the texture motion. 
The motion of our temporal texture can be quantified by comparing the variation of rendered views in adjacent time frames with camera pose $\Phi$.
Specifically, let $P_b$ and $P_f$ be two rendered images of the stylized density field with VNCA from $\Phi$ during training, with $P_b$ as the image rendered before a random $n$ update steps is applied to VNCA, and $P_f$ as the image after the updates. 
We use a pre-trained optical flow prediction network $\mathcal{F}_\text{OF}$~\cite{tesfaldet2018two} to predict the perceived motion between $P_b$ and $P_f$ given by $\mathcal{J}^r = \mathcal{F}_\text{OF}(P_b, P_f)$ as 2D vector. 
We align the directions of the target vector field and the predicted vector by computing the cosine distance given by$\mathcal{L}_{dir}$:
\begin{equation}
\mathcal{L}_\text{dir} = \frac{1}{H\cdot W} \sum_{i, j}d_c(\mathcal{J}^r _{ij}, \mathcal{J}^\psi _{ij}). 
\end{equation}

To obtain a texture motion that has roughly the same magnitude as the input vector field, we normalize the predicted vector field with the number of update steps $n$ between $P_b$ and $P_f$ and the $N$ update steps we use between each pair of adjacent rendered views:
\begin{equation}
\mathcal{L}_\text{mag} = \frac{1}{H\cdot W} \sum_{i, j}  \left | \frac{N}{n} \cdot  \norm{\mathcal{J}^r_{ij}} _2 -\norm{\mathcal{J}^\psi_{ij}}_2 \; \right |.
\label{eq:norm_loss}
\end{equation}

The final motion supervision loss $\mathcal{L}_{motion}$ is given by:
\begin{equation}
\mathcal{L}_\text{motion} =  \max(0, \mathcal{L}_\text{dir} - 1) \cdot \lambda_\text{mag}\mathcal{L}_\text{mag} + \lambda_\text{dir} \mathcal{L}_\text{dir},
\end{equation} 
such that the training prioritizes direction alignment between the target and predicted motion over their magnitudes.

At each epoch, we condition one density field to ensure that the optical flow network is not biased by the motion of the input density sequence. We rotate the camera position between epochs for multiview flow-guided supervision of texture motion. 
For inference, we apply $N$ update steps on the cell states for a visible motion on each pair of adjacent density fields.

%% file: sec/4-exp.tex
\begin{figure}
  \centering
  \includegraphics[width=0.9\linewidth]{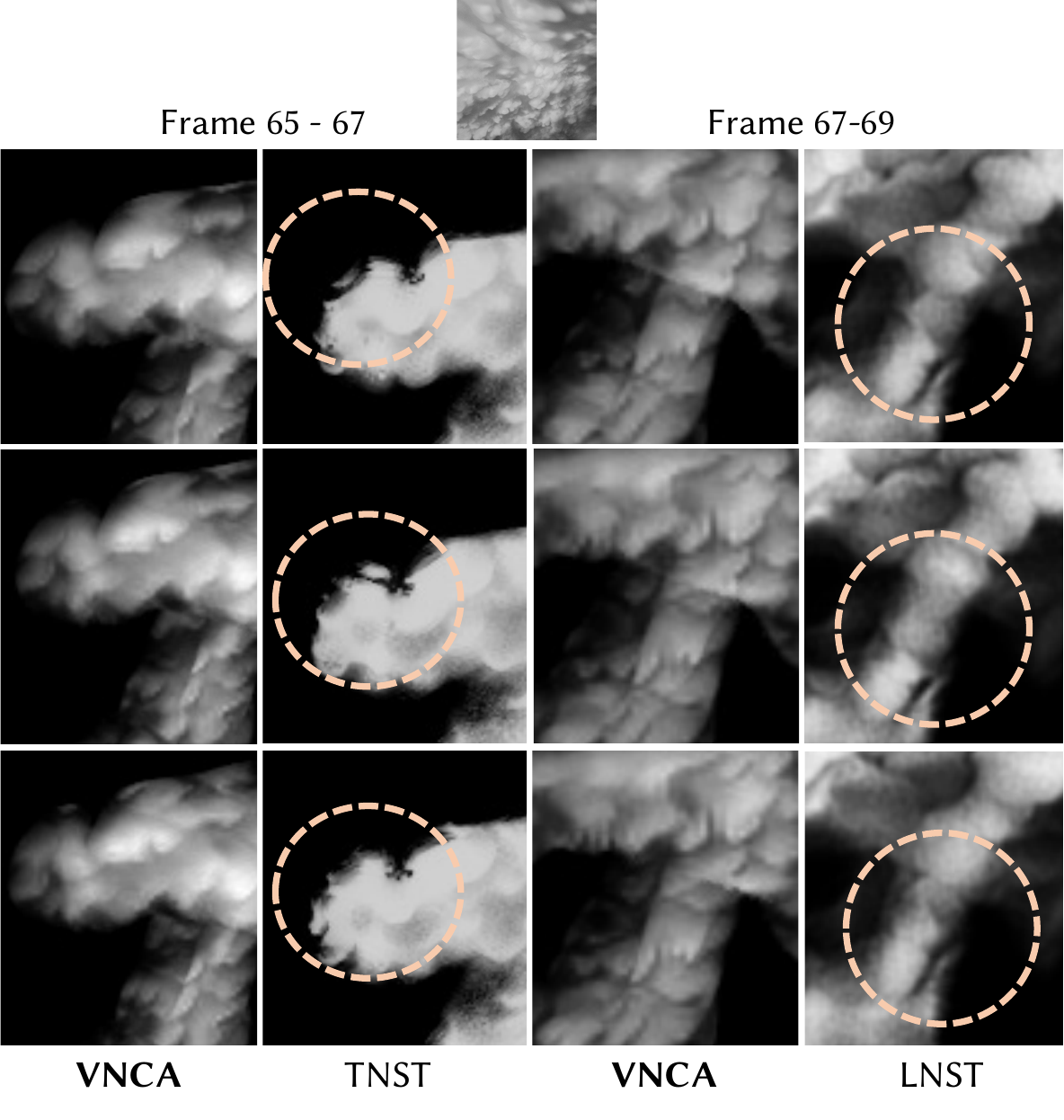}
  \caption{\textbf{Between-frame Coherence Comparison with TNST and LNST.} VNCA synthesizes more coherent stylization across frames, compared to per-frame optimization methods. }
  \label{fig:test}
\end{figure}

\input{sec/table-stat}

\section{Experiments}

\subsection{Qualitative Results} 
To demonstrate that our method works for a variety of style images, we show the stylization results on three smoke data \textit{Smoke Jet}, \textit{Bunny}, and \textit{Billowy} in \Cref{fig:res}. 
VNCA can synthesize RGB colors for the input density sequence and modify the input density field to match the style image, as shown on the boundaries of each volume dataset that matches the pattern given by the style image.  
Our method supports different style images and can capture the key features in the reference image for aesthetic stylization. 

In the stylization of \textit{Mammatus} and \textit{Wave} on \textit{Billowy}, the synthesized color is brighter at the bottom of the simulation. 
We empirically find that regions with more concentrated density lead to brighter synthesized style colors.
By changing the transmittance of the density field, we can exhibit control over the brightness of stylized smoke without additional training.

\subsection{Comparison with Prior Works}
\paragraph{Baseline Methods}
To our knowledge, there is no other work on the same problem, i.e. volumetric temporal texture synthesis to stylize 3D smoke data. We thus compare VNCA with 3D smoke stylization methods \citet{kim19c} and \citet{kimlnst}.

\begin{figure}
  \centering
  \includegraphics[width=0.9\linewidth]{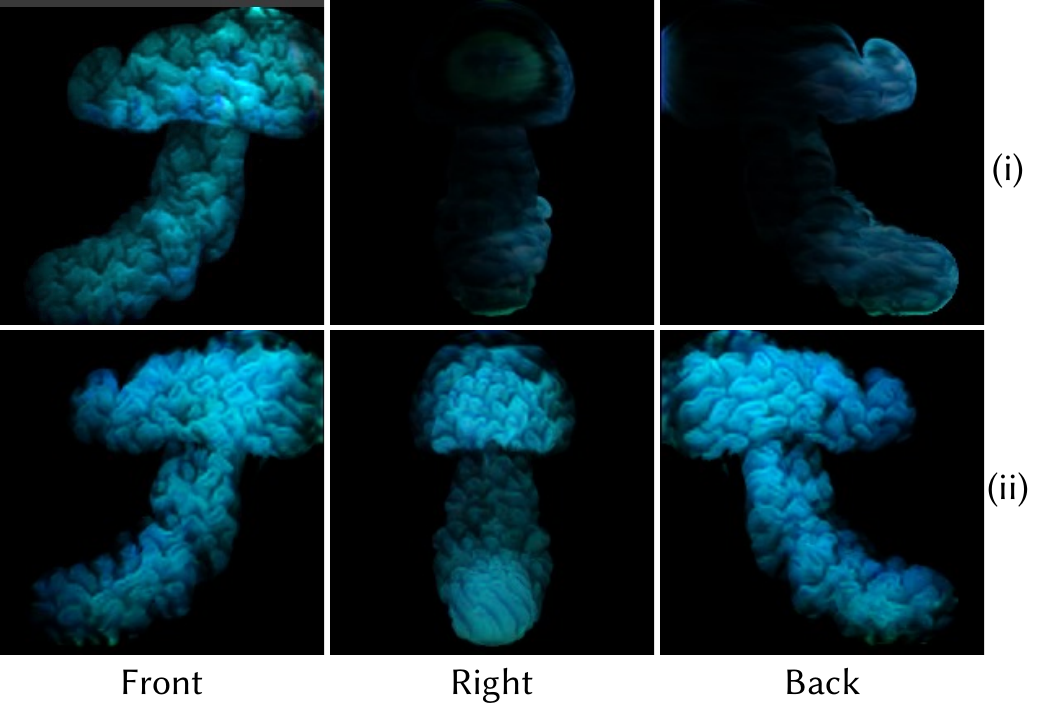}
  \caption{\textbf{Ablation Studies on Multiview Consistency. } (i) front view supervision; (ii) randomized view supervision. If VNCA conditions on a single view during training, we cannot achieve omniview consistent stylization. }
  \label{fig:multiview}
\end{figure}

\begin{figure}
  \centering
  \includegraphics[width=0.9\linewidth]{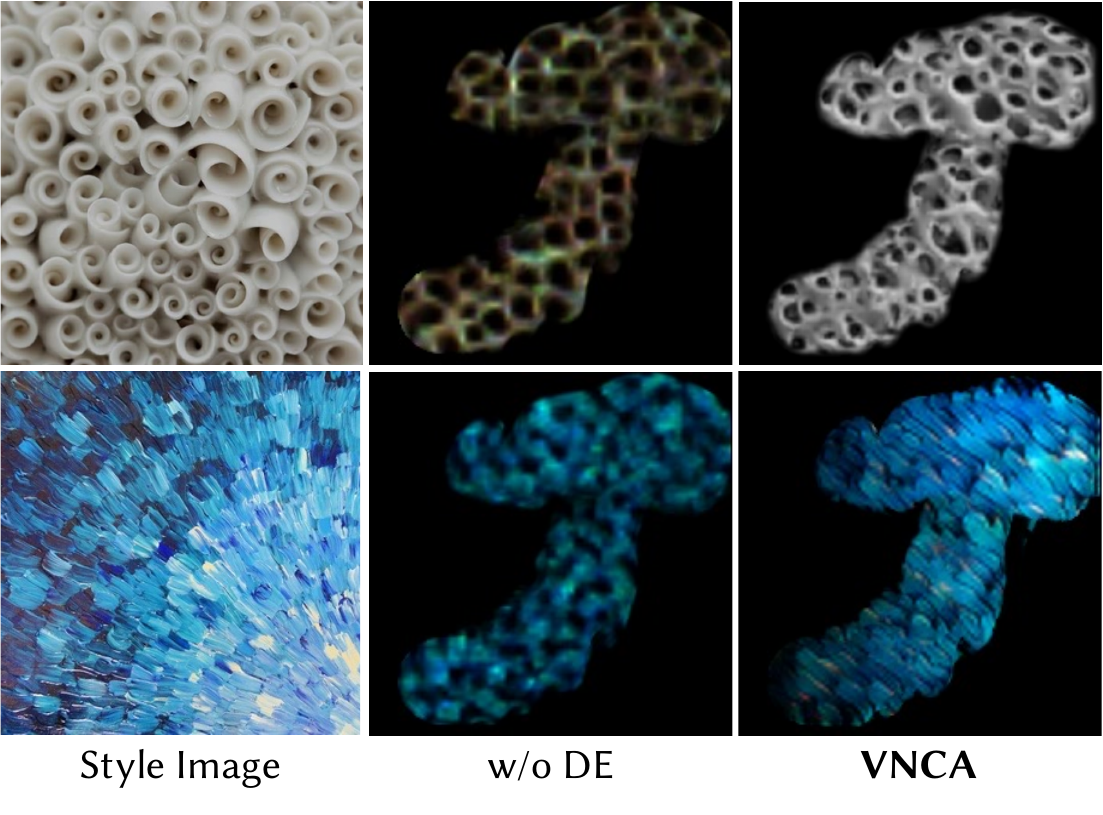}
  \caption{\textbf{Ablation Studies on Density Encoding. } Without density encoding, the synthesized appearance is porous and inconsistent with input reference. With density encoding, the output is much improved. }
  \label{fig:de}
\end{figure}

\paragraph{Stylization Comparison}
Our work focuses on 3D volumetric temporal texture. 
Since TNST and LNST do not support color stylization on \textbf{3D smoke data}, we use grayscale reference images for all methods. 
We run experiments on the same RTX3090Ti for fairness. 
\Cref{fig:comp} shows the qualitative comparison of stylization matching quality. VNCA can synthesize features that match the reference pattern and turbulence effect on par with TNST and LNST.

\Cref{fig:test} compares the preservation of stylization coherence between frames. 
We highlight the circled regions for inconsistent stylization between consecutive frames in prior works,
as they stylize each density field independently, inevitably leading to inter-frame inconsistencies. 
In contrast, the volumetric temporal texture synthesized by VNCA is naturally dynamic from the emerging motion of the underlying NCA model.

\Cref{tab:stat} shows the results of our user study on the stylization quality, inter-frame consistency, and preservation of smoke motion comparison for TNST, LNST, and VNCA. 
We received 50 valid user responses for stylization quality evaluation ("Which image has the texture pattern that best matches the reference image?"), 44 for inter-frame consistency ("Which video has a smoother transition between frames?"),
and 49 for motion preservation evaluation ("Which video is most plausible as a natural smoke sequence?"). 
TNST is not compared for inter-frame coherence as LNST is shown to be better~\cite{kimlnst}. 
Each entry shows the percentage of users that prefer the stylization of corresponding methods. 
We see a preference for our methods over the baseline methods.

\paragraph{Runtime Comparison}
In \Cref{tab:stat}, we provide a quantitative comparison with TNST and LNST on the time required to stylize one frame of the input sequence. 
We measure this data by stylizing 20 consecutive frames and taking the average time to take into account the inter-frame smoothing time which impedes the performance of TNST and LNST. 
We can see that our average inference time for a single frame is faster than prior methods, while our training already includes omni-view consistency.

\begin{figure}
  \centering
  \includegraphics[width=0.9\linewidth]{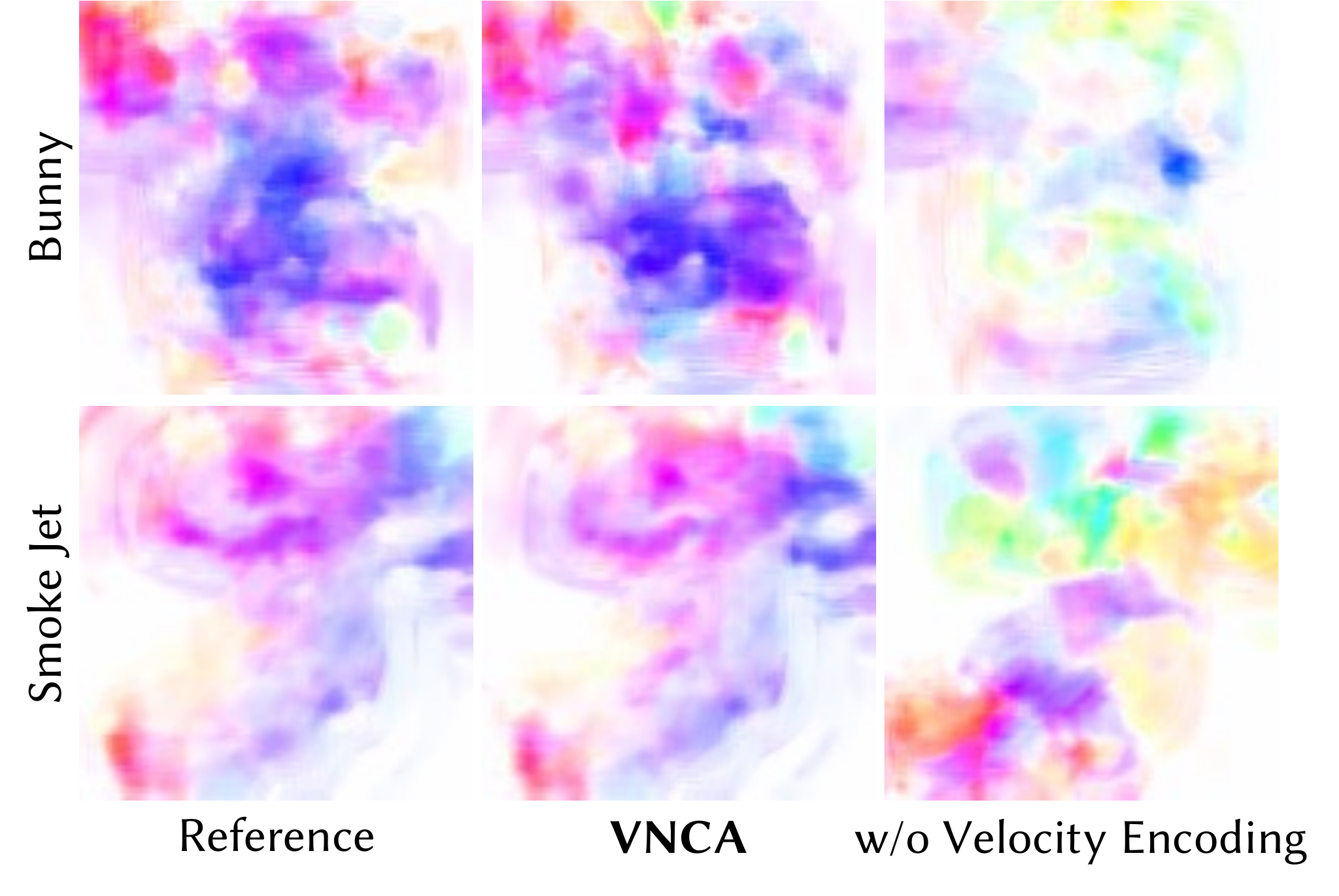}
  \caption{\textbf{Ablation Studies on Velocity Encoding.} The use of velocity encoding better aligns the estimated optical flow of our synthesized texture motion with the input velocity field. }
  \label{fig:OF}
\end{figure}

\subsection{Ablation Studies}
\label{sec:abl}
We hereby evaluate the design choice of VNCA on encoding priors and training strategies in our ablation studies, and we further demonstrate that VNCA is not a trivial adaptation of prior works. 

\paragraph{Multi-view Style Consistency. } 
During training, we randomly select one of the camera viewpoints to render and compare it with the reference style image. 
This approach ensures that the stylized volume contains consistent style features from different viewpoints. 
\Cref{fig:multiview} shows the visual comparison of rendered viewpoints from the front, right (front $+90\degree$), and back (front $+180\degree$) between a stylized volume supervised by (i) only frontal view, and (ii) from a randomly selected viewpoint. 
We can see that only conditioning on one viewpoint cannot generalize to omni-view style consistency, demonstrating the importance of our multi-view supervision. 

\paragraph{Density encoding.} 
We conduct an ablation study on the density encoding in the Volumetric NCA framework. 
By appending one of the input density frames to the perception vector, we provide our NCA model with information about the global structure of the density distribution in the input data. 
This global information is crucial for appearance stylization since each cell in the NCA's cell state is only aware of its immediate neighborhood. 
As shown in~\Cref{fig:de}, the stylization without density encoding results in an overall color that is somewhat close to the reference but is abundantly porous and lacks detailed structures, such as stroke patterns.
Including density encoding significantly enhances the quality of stylization in terms of matching the given style image.

\paragraph{Velocity encoding} 
The use of velocity encoding facilitates the alignment of our volumetric temporal texture and guiding velocity field. 
We compare the estimated optical flow between the synthesized texture motion and the projected velocity field on the smoke datasets in \Cref{fig:OF}. 
We fix the density frame to not bias the network with fluid motion. 
The synthesized result with velocity encoding shows a closer alignment to the guiding motion flow. 

\Cref{tab:ablation} shows a quantitative comparison of the impact of velocity encoding on the alignment of motion direction and magnitude, and the impact of density encoding on the appearance supervision.

\begin{figure}
  \centering
  \includegraphics[width=\linewidth]{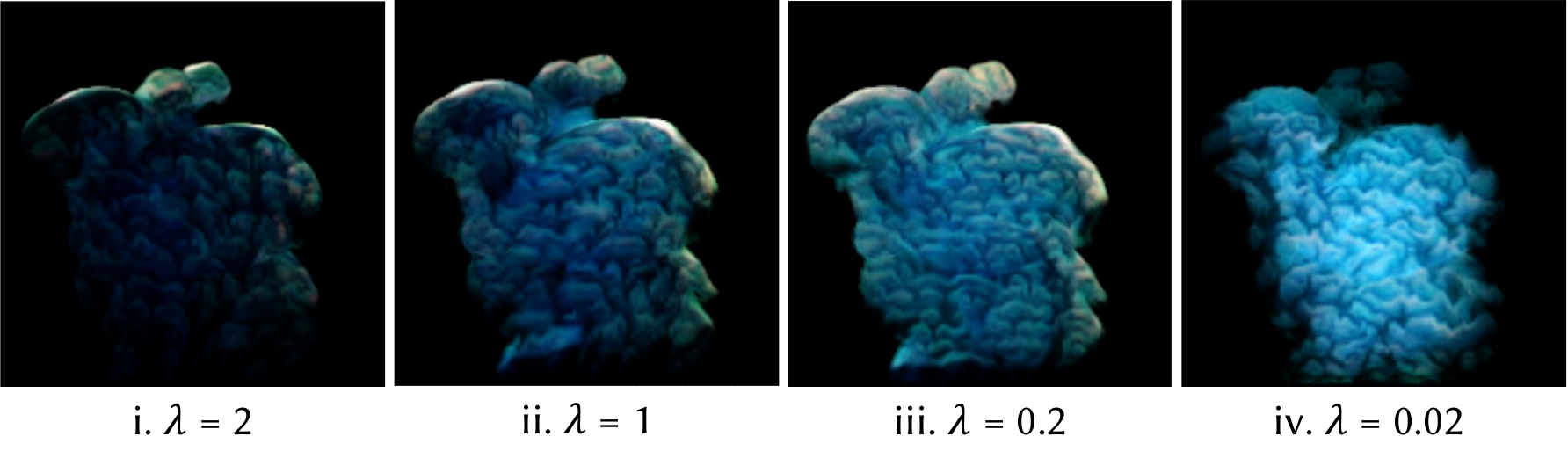}
  \caption{\textbf{Comparison of different weights for loss functions. } $\lambda = \lambda_\text{motion} / \lambda_\text{app}$. When the weight for the motion loss is too high w.r.t appearance loss, the stylization degenerates. (Style Image: \textit{Ink})  }
  \label{fig:param}
\end{figure}

\paragraph{Comparison on weights for loss functions} \Cref{fig:param} shows the ratio $\lambda = \lambda_\text{motion}/\lambda_\text{app}$ between the weight of $\mathcal{L}_\text{motion}$ to that of $\mathcal{L}_\text{app}$. 
The stylized appearance can be corrupted when the weight for the motion loss is too high compared to the appearance loss.

\input{sec/table-ablation}

\subsection{Generalization Ability}
\label{sec:gen}
A key feature of VNCA stems from the robust generalizability of NCA models. 
At each epoch, we condition VNCA on the current density field of the training frame. 
This allows the synthesized volumetric temporal texture to generalize to other unseen density fields in the input smoke sequence at inference time.
As we condition the density field to be stylized during inference and supply the velocity field, VNCA does not require re-training to synthesize matching appearance and corresponding visually plausible motion.

In this way, a converged volumetric temporal texture can stylize the entire density sequence. We empirically find that trained VNCA can even extrapolate to smoke datasets unseen at training time. 

\begin{figure}
  \centering
  \includegraphics[width=\linewidth]{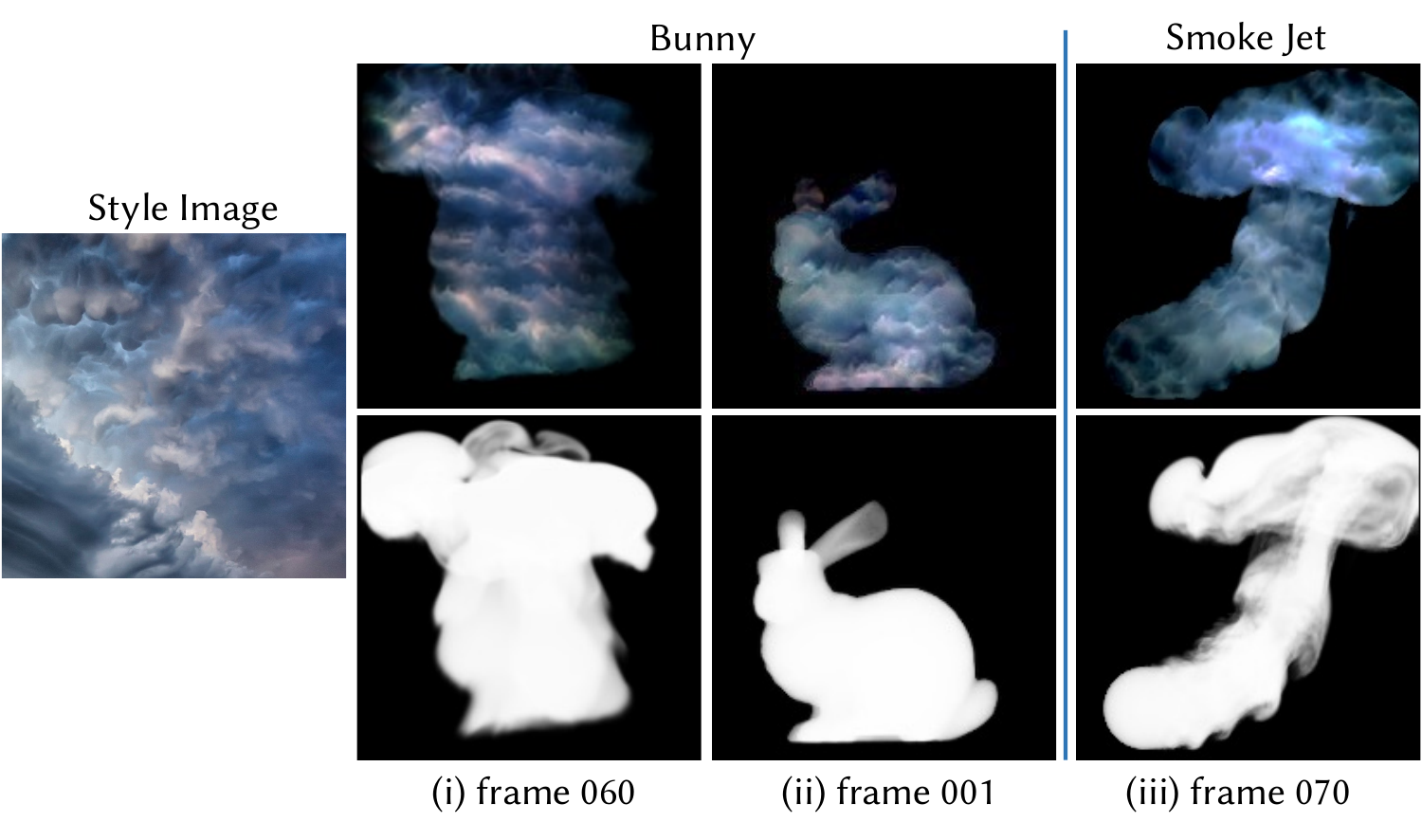}
  \caption{\textbf{Generalization Ability of VNCA. } During training, we condition VNCA on the 60th frame of \textit{Bunny}. At inference, the trained VNCA can stylize the full sequence of \textit{Bunny}, and \textit{Smoke Jet} which is unseen in training.  }
  \label{fig:gen}
\end{figure}

In \Cref{fig:gen}, we illustrate the generalizability of our volumetric temporal texture. The converged VNCA is conditioned on the 60th frame of the \textit{bunny} dataset in (i). 
At inference time (ii), the trained texture volume can stylize not only the 60th frame but also all other frames of the bunny scene.
Additionally, (iii) it is capable of stylizing a new dataset \textit{Smoke Jet} unseen during training, as demonstrated by the results on the 70th frame. 
This versatility allows our model to reduce training time compared to baseline methods.

%% file: sec/table-stat.tex
\begin{table}
\scalebox{0.9}{
\begin{tabular}{ccccccc}
\hline
Method  &  \small{Runtime}              & \small{Stylization}  &  \small{Inter-frame} & \small{Motion}\\
       &  &   \small{Quality}    & \small{Consistency} & \small{Preservation}\\
\hline
TNST           & 463s      &  0.67\%    &  n/a      &  16.33\% \\
LNST           & 10s       &  19.33\%   &  15.91\%  &  28.57\% \\
\textbf{VNCA}  & 0.8s      &  80\%      &  84.09\%  &  55.10\%
\tabularnewline \hline
\end{tabular}
}
\caption{\textbf{Quantitative Comparison with TNST and LNST. } We quantitatively compare the methods on the \textit{Smoke Jet} dataset ($200\times300\times200$) with publically available code on the required per-frame stylization time and conduct a user study to compare the Inter-frame consistency and stylization quality. The runtime for prior works is evaluated \textbf{without omni-view training} on an RTX3090Ti. }
\label{tab:stat}
\end{table}

%% file: sec/table-ablation.tex
\begin{table}

\scalebox{0.9}{
\begin{tabular}{c|cc|c|cc}
\hline
Loss   & \small{Positional}  & \small{\textbf{P + V}}   & Loss  & \small{w/o Density} &  \small{\textbf{Density}} \\
       &\small{Encoding}     &\small{\textbf{Encoding}} & & \small{Encoding} &\small{\textbf{Encoding}} \\
\hline
$\mathcal{L}_\text{dir}$  & 0.134     & \textbf{0.083}     & $\mathcal{L}_\text{app}$ 
 & 7.29 & \textbf{5.42} \\
$\mathcal{L}_\text{mag}$  & 0.295     &  \textbf{0.167}    &  -  & 
\tabularnewline \hline
\end{tabular}
}
\caption{\textbf{Quantitative Ablation on the Encoded Priors.} By encoding the density field and velocity field of the smoke sequence, the synthesized temporal texture of VNCA achieves better appearance and motion alignment with the stylizing target.   }
\label{tab:ablation}
\end{table}

%% file: sec/5-mesh.tex
\section{Application to Mesh Texturing}
\label{sec:mesh_texturing}
VNCA can easily be extended for mesh texturing given a reference style following the line of research in Solid Texture Synthesis \cite{perlin_noise, kopf2007solid, on-demand-solid-texture}.
\Cref{fig:mesh} shows the stylized meshes by our method in comparison with the results from A. \citet{on-demand-solid-texture} and B. \citet{kopf2007solid} as two representative solid texturing methods for five different style images.

%% file: sec/6-concl.tex
\section{Conclusion and Discussion}

In this paper, we have presented VNCA, a novel volumetric temporal texture for stylizing volumetric smoke sequences with 2D exemplars. Our method achieves 1. high-quality appearance stylization with multiview and across-frame coherence w.r.t a reference image, 2. visually plausible motion as the original smoke data, and 3. speedup training time.
Our model can be extended to mesh texturing with a texture volume to be applied on various meshes.

\subsection{Limitations. }
Since we use the NCA framework, we inherit its shortcomings in that the homogenous update rule for NCA limits the synthesized texture from achieving high heterogeneity. 
In other words, VNCA can extract repeated patterns in the style image as features for stylization, but may not be able to synthesize stylizing features with a clear foreground and background.